\documentclass[aps,pre,reprint,groupedaddress]{revtex4-2}

\usepackage{lineno,hyperref,amsmath,graphicx}
\graphicspath{ {figures/} }

\begin{document}


\title{Time-Dependent Density Functional Theory Applied to Average Atom Opacity}


\author{N. M. Gill}
\email{ngill@lanl.gov}

\author{C. J. Fontes}
\author{C. E. Starrett}

\address{Los Alamos National Laboratory, P.O. Box 1663, Los Alamos, NM, 87545, USA}

\date{\today}

\begin{abstract}
We focus on studying the opacity of iron, chromium, and nickel plasmas at conditions relevant to experiments carried out at Sandia National Laboratories [J. E. Bailey \textit{et al.}, Nature \textbf{517}, 56 (2015)]. We calculate the photo-absorption cross-sections and subsequent opacity for plasmas using linear response time-dependent density functional theory (TD-DFT). Our results indicate that the physics of channel mixing accounted for in linear response TD-DFT leads to an increase in the opacity in the bound-free quasi-continuum, where the Sandia experiments indicate that models under-predict iron opacity. However, the increase seen in our calculations is only in the range of 5--10$\%$. Further, we do not see any change in this trend for chromium and nickel. This behavior indicates that channel mixing effects do not explain the trends in opacity observed in the Sandia experiments.
\end{abstract}


\maketitle

\section{Introduction}

Experiments carried out at Sandia National Laboratories obtained results for iron opacity that differed significantly from commonly used opacity models at some plasma conditions but not for others \cite{bailey2007, bailey2015}. Similar experiments were carried out for nickel and chromium, with similar discrepancies for certain plasma conditions in chromium observed but not in nickel \cite{nagayama2019}. This controversy has led to a surge of research to determine whether the missing opacity is due to physics lacking in theoretical models or due to experimental error (for example, see the following references \cite{colgan2016, nahar2016, mancini2016, more2017, gill2019}). Beyond the specific application to the Sandia experiments, opacity is a key quantity in the study of stellar interiors \cite{serenelli2009, salmon2012}, white dwarf simulations \cite{piron2018}, and simulations of inertial confinement fusion experiments \cite{hu2014_opac}.

Density functional theory (DFT) provides a convenient theoretical framework to study dense systems by using the electron density as the central quantity of interest. DFT models have proven widely successful in studying many-electron systems, but DFT models only give information on the thermal ground state of system and are therefore poorly suited to studying spectral quantities that require knowledge of excited states of the system. The generalization of DFT to include excited states was accomplished with time-dependent density functional theory (TD-DFT). The TD-DFT formalism was, in large part, laid out in work done by Zangwill and Soven (ZS), where they applied to the concept of a dynamic response of electrons to a perturbation in the study of photo-absorption in cold neutral gases \cite{zangwill80}. Throughout the 1980s, ZS's approach was utilized to study polarizability in many-particle systems (for a good description of the theory as it developed in the 1980s, see the text by Mahan and Subbaswamy \cite{m_and_s_1990}). The ZS formalism inspired a rigorous formulation of TD-DFT \cite{runge84, bartolotti82, vanLeeuwen99, marques2004}.

Average atom models are usually DFT-based models of atoms embedded in plasmas, and they have seen widespread use in the study of thermodynamic quantities in dense plasmas (for example, see the following references \cite{liberman, piron11, piron2018, starrett2019}). Most average atom models have the benefit of implicitly accounting for plasma density effects through the use of a finite-sized spherical cavity in which all of an ion's electrons and the nucleus are contained. Before the formalization of TD-DFT, Grimaldi et al. applied the techniques of ZS to finite-temperature plasmas using an average atom model to determine the ground state of their system \cite{grimauldi85}. Their results for polarizability in plasmas showed resonances arising from interference between different absorption processes. Such effects are missing from independent particle treatments for opacity \cite{hu2014_opac, shaffer, johnson2020, faussurier2018, desjarlais2005, ovechkin2014}.

In this work we apply TD-DFT under the linear response formalism to study the opacity of an average atom. Our average atom model is calculated using the \texttt{Tartarus} code \cite{gill17, starrett2019}, and the linear response formalism of finite-temperature TD-DFT that we apply is similar to that implemented by Grimaldi et al \cite{grimauldi85}. The focus of our results is to study the opacity of plasmas at conditions relevant to the aforementioned Sandia experiments. The goal is to demonstrate the effects of interference between different absorption processes in the plasma---the so-called channel mixing effect. Due to our use of an average atom model to describe the plasma as well as our use of DFT for the solution of the electronic structure, our bound-bound opacity is not quantitative. Thus the focus of our work is to show the importance of the channel mixing effect on the bound-free features of the opacity.

In the remainder of this paper, we first provide an outline of the theoretical formalism used in our calculations. Next, some of the calculation details are described, including approximations used and the level widths that were implemented in our models to overcome poles in the density response functions. Finally, we present the results of our calculations, starting with verification of the code by comparing with results previously obtained by other authors. We also show opacities for iron, chromium, and nickel plasmas near the reported conditions of the Sandia Z-machine experiments. We discuss the relative impact of the channel mixing effect and show that, although the redistribution of the various spectral contributions does lead to an increase in bound-free tails of the opacities, channel mixing accounted for at the level of linear response does not explain the so-called ``missing opacity'' indicated by the experiments. An additional discussion on the dynamic response of free electrons is included in Appendix A. 

\section{Theoretical Formalism}
\subsection{Linear Response TD-DFT}
The following section briefly summarizes the linear response formalism of TD-DFT, with a prescription that is commonly found in other work \cite{zangwill80,m_and_s_1990,marques2004}. We consider a system in its ground state, described by the electron density, $n_{0}(\vec{r})$. At time $t=0$, the system becomes perturbed by a weak electric field, leading to an evolution of the density over time, i.e. the electron density now has time dependence, $n(\vec{r},t)$. Throughout this paper, Hartree atomic units are utilized ($\hbar=m_{e}=e=1$ and energy is labeled as $\mathrm{E_{H}}$). 

The total external potential seen by the electrons includes the ground state external potential and the perturbing potential,
\begin{equation}
\label{total_v}
V(\vec{r},t) = V_{ext}(\vec{r},0) + \delta V_{pert}(\vec{r},t)
\end{equation}
where $V_{ext}(\vec{r},0)$ is the ground state external potential and $\delta V_{pert}$ is the potential associated with a small perturbation. From response theory, we then consider an expansion of the density with respect to the perturbing potential about the ground state density:

\begin{equation}
\label{density_expansion}
n(\vec{r},t) = n_{0}(\vec{r}) + \delta n(\vec{r},t) + \cdots
\end{equation} 
where the first-order perturbed density is given by

\begin{equation}
\label{perturbed_dens_resp}
\delta n(\vec{r},t) = \int_{0}^{\infty} dt' \int_{V_{ion}} d\vec{r'} \delta V_{pert}(\vec{r'},t') \chi(\vec{r},\vec{r'},t,t')
\end{equation}
The quantity $\chi$ is the linear response function of the electrons. If a Fourier transform is carried out, we can look at the frequency representation of the quantities of interest, which gives us the frequency-dependent induced electron density,
\begin{equation}
\label{ind_n_omega}
\delta n(\vec{r},\omega) = \int d\vec{r'} \delta V_{pert}(\vec{r'},\omega)\chi(\vec{r},\vec{r'},\omega)
\end{equation}
where $\omega$ is the frequency associated with the perturbing photon.

Though $\chi$ is not known \textit{a priori}, we can determine its form through a self-consistent procedure. In terms of the independent particle Kohn-Sham (KS) response function, $\chi_{0}$, the interacting linear response function can be written as \cite{marques2004}
\begin{equation}
\label{scf_chi_full}
\begin{split}
&\chi(\vec{r},\vec{r'},\omega) =  \chi_{0}(\vec{r},\vec{r'},\omega) + \\  &\int d\vec{r_{1}} \chi_{0}(\vec{r},\vec{r_{1}},\omega) \int d\vec{r_{2}} \Bigg [ \frac{1}{|\vec{r'}-\vec{r_{2}}|} + \frac{\delta V_{xc}(\vec{r'},\omega)}{\delta n(\vec{r_{2}},\omega)}\Bigg |_{n_{0}} \Bigg ] \chi(\vec{r_{2}},\vec{r_{1}},\omega)
\end{split}
\end{equation}
where $V_{xc}(\vec{r'},\omega)$ is the Fourier transform of the time-dependent exchange-correlation potential. The terms in brackets represent the Coulomb interaction and exchange-correlation interactions between electrons in the system. Equation \ref{scf_chi_full} can be solved self-consistently by using $\chi_{0}$ as an initial guess for $\chi$ on the right-hand side. The function $\chi_{0}$ is fully determined by
\begin{equation}
\label{ind_chi}
\begin{split}
\chi_{0}(\vec{r},\vec{r'},\omega) = \sum_{i,j} (f_{i}-f_{j})\frac{\psi_{i}^{*}(\vec{r})\psi_{j}(\vec{r})\psi_{i}(\vec{r'})\psi_{j}^{*}(\vec{r'})} {\omega - (E_{j}-E_{i}) + i\delta}
\end{split}
\end{equation}
where the $\psi_{i}$ are the KS single-particle orbitals of energy $E_{i}$ and occupation $f_{i}$, and $\delta$ is a positive infinitesimal quantity. It is important to note that the sum in equation \ref{ind_chi} runs over all possible KS states of the system, including states of positive energy, and therefore its evaluation would require accounting for an infinite number of states when determining $\chi_{0}$. The Green's function associated with the KS Hamiltonian, $G$, can be employed to simplify equation \ref{ind_chi} and implicitly account for one of the needed sums over all states. The following expression was shown by ZS and is the practical expression utilized to calculate linear response functions:
\begin{equation}
\label{ind_chi_gf}
\begin{split}
\chi_{0}(\vec{r},\vec{r'},\omega) =& \sum_{i} f_{i}\psi_{i}^{*}(\vec{r})\psi_{i}(\vec{r'})G(\vec{r},\vec{r'},E_{i}+\omega) \\+ &\sum_{i}f_{i}\psi_{i}(\vec{r})\psi_{i}^{*}(\vec{r'})G^{*}(\vec{r},\vec{r'},E_{i}-\omega)
\end{split}
\end{equation}

\subsection{Opacity}
The absorption cross-section is related to the induced density by
\begin{equation}
\label{pa_xsection}
\sigma(\omega) = 4\pi \frac{\omega}{c} \Im {\alpha(\omega)}
\end{equation}
where $c$ is the speed of light, $\Im$ indicates taking the imaginary part, and $\alpha$ is the polarizability, which is given by
\begin{equation}
\label{polar}
\alpha(\omega) = \frac{-2}{\varepsilon_{0}}\int d\vec{r} \, z \, \delta n(\vec{r},\omega)
\end{equation}
where $\varepsilon_{0}$ is the magnitude of the electric field associated with the perturbation.

The opacity can then easily be determined from the photo-absorption cross-section by
\begin{equation}
\kappa(\omega) = \frac{\sigma(\omega) n_{ions}}{\rho}
\end{equation}
where $n_{ions}$ is the number density of ions in the system and $\rho$ is the mass density of the plasma.

\section{Calculation Details}
\subsection{Ground State Calculation}
We employ the \texttt{Tartarus} average atom model in order to calculate our ground state electron density. Here we will provide a brief overview of the model. For more details on the calculation, we defer the reader to reference \cite{starrett2019}. The \texttt{Tartarus} average atom model consists of an atom in a sphere of volume $V_{ion}$, where $V_{ion}$ is the average volume taken up by ions in the plasma. The atom is enforced to be charge neutral with its electronic structure determined using KS DFT. The key approximation used in KS DFT is the choice of an exchange-correlation (XC) functional, for which a local density approximation (LDA) form of XC functional is chosen in \texttt{Tartarus}. The plasma screening is such that all external fields are completely screened beyond the ion sphere, enforcing a zero effective potential at the sphere boundary and beyond. 

The electrons and nucleus are assumed to be in local thermodynamic equilibrium, and Fermi-Dirac statistics are applied to the occupations of electronic states, i.e. $f_{i} \rightarrow f(E_{i},\mu)$, where $f$ is the Fermi-Dirac function and $\mu$ is the chemical potential of the system. The solution of the \texttt{Tartarus} model yields the electronic chemical potential of the plasma and an effective potential (which corresponds to a complete set of KS eigenstates and an associated electron density). Though \texttt{Tartarus} can generate both relativistic \cite{gill17} and nonrelativistic \cite{starrett15} solutions, we only utilize the nonrelativistic average atom ground state in this work.

\subsection{TD-DFT Implementation}
In this work, we implement the linear response TD-DFT formalism using the \texttt{Tartarus} ground state density. Though equations \ref{scf_chi_full} and \ref{ind_chi_gf} contain the necessary information needed to carry out a self-consistent determination of $\chi$, it is algorithmically simpler to decompose the self-consistent procedure in terms of the induced electron density, $\delta n$, and an effective induced potential, $\delta V_{ind}$. 

Using KS DFT we define an effective potential, $\delta V_{KS}$, such that the induced electron density can be determined via the KS independent response function:
\begin{equation}
\label{dn_ind_eff}
\delta n(\vec{r},\omega) = \int d\vec{r'} \chi_{0}(\vec{r},\vec{r'},\omega) \delta V_{KS}(\vec{r'},\omega)
\end{equation} 
The effective potential contains the potential associated with the perturbation (in our case the dipole potential associated with the incident photon) and the induced potential that arose from the response of the electrons to each other while under the influence of the perturbation, i.e. 

\begin{equation}
\label{KS_eff_V}
\delta V_{KS}(\vec{r},\omega) = \delta V_{pert}(\vec{r}) + \delta V_{ind}(\vec{r},\omega)
\end{equation}
where $\delta V_{ind}$ is the induced potential arising from the electrons' interactions with each other. $\delta V_{ind}$ represents the effective potential seen by each KS electron after it is perturbed from the ground state and is given by
\begin{equation}
\label{V_ind_label}
\delta V_{ind}(\vec{r},\omega) = \int d\vec{r'} \frac{\delta n(\vec{r'},\omega)}{|\vec{r}-\vec{r'}|} + \delta V_{xc}(\vec{r},\omega)
\end{equation}

Equations \ref{dn_ind_eff} and \ref{KS_eff_V} can be solved self-consistently with the initial iteration using $\delta V_{ind}=0$. The first iteration yields the so-called independent particle induced density, $\delta n_{0}$. This quantity can be used to determine the response of the ground state system without allowing the electrons to rearrange in response to each other. Once iterated to convergence, the induced density obtained using equation \ref{dn_ind_eff} is, to some numerical tolerance, equivalent to that obtained via equation \ref{ind_n_omega}.

The equations presented so far for the KS TD-DFT formalism have not been written with any specific form of the exchange-correlation functional. In principle, detailed KS TD-DFT calculations require a time-dependent exchange-correlation functional, but efforts to develop such functionals with successful application across a variety of systems is an ongoing area of research (see, for example, references \cite{vignale1996, tokatly2009, kurzweil2004, wijewardane2008} and their citing articles). In lieu of building in explicit time (or frequency) dependence into the exchange-correlation functional, most modern applications use the adiabatic local density approximation (ALDA) \cite{marques2004}. In this approximation, the exchange-correlation functional is a linearization about the ground state electron density of a time-independent LDA exchange-correlation functional:
\begin{equation}
\label{alda}
\delta V_{xc}(\vec{r},\omega) = \delta n(\vec{r},\omega) \frac{\delta V_{xc,0}(\vec{r})}{\delta n(\vec{r'})}\Bigg |_{n=n_{0}}
\end{equation}
where $V_{xc,0}$ is the exchange-correlation potential at $t=0$, i.e. the ground state exchange-correlation potential.

From equation \ref{alda} it is clear that the only frequency dependence that enters into the ALDA comes from the frequency dependence of the induced electron density. This approximation is expected to perform well when studying perturbations that occur over small time scales in finite-sized systems and finds wide usage in the study of the optical properties of matter (see, for example, reference \cite{octopus2020}). All of the calculations demonstrated in this work were performed under the ALDA, with the use of Dirac's form of the exchange-correlation energy functional \cite{dirac}.

\section{Numerical Details}
\subsection{Level Width}
One common difficulty in the self-consistent calculation outlined by equations \ref{dn_ind_eff} and \ref{KS_eff_V} is the numerical instability in the real part of $\delta n$ at frequencies near the poles of $\chi_{0}$. These poles occur only in the imaginary part of $\chi_{0}$ at the bound-bound excitation energies of the KS ground state, but the real part of $\chi_{0}$ is also large enough near these poles to create numerical instability in the self-consistent procedure, which can lead to a false or failed convergence.

Careful treatment of these large values is necessary to successfully reproduce Fano resonances when a bound-bound absorption process interferes with a continuum absorption process. In order to stabilize the calculation, an imaginary factor can be added to KS state energies, but this introduces a free and adjustable parameter into the calculation that could introduce unphysical trends into results. Grimaldi et al. \cite{grimauldi85} utilized a self-energy correction in order to obtain physically meaningful values for this imaginary factor, which we use in our calculations. 

These corrections are generally quite small, and we also implement a simple Stark broadening scheme which serves a similar purpose while also giving broadened bound-bound features in the resulting spectra. The level widths are related to the Stark shift in energies given by a simple first-order perturbation theory calculation using the Holtsmark distribution as an approximation of the electric fields in the plasma. This level width is generally larger than that obtained via the self-energy correction but is still quite small except in the case of very weakly bound states.

Since our physical description of the plasma comes from an average atom model, our electronic structure corresponds to a single, averaged ion stage. As such, we do not have explicit inclusion of multiple ion stages or electronic configurations, which means the bound-bound features in our opacity are not quantitatively comparable to realistic spectra. This approach allows us to use these simple level width schemes, as the bound-bound opacity resulting from our calculation only serves as a qualitative description of the many bound-bound absorption processes present in an actual plasma. We also note that the use of the level width models mentioned here has a negligible effect on the bound-free features of the spectra.

\subsection{Dynamic Response of Continuum Electrons}
A well-known problem when studying photoexcitation between two continuum electrons is the divergence of the dipole matrix elements needed to calculate a cross-section \cite{johnson06, kuchiev}. Though the linear response TD-DFT approach does not involve explicit use of such matrix elements, the same issue arises implicitly in the determination of $\delta n$. Although this problem has been addressed in work by others \cite{blenski2006, caizergues16}, treating free-free transitions with a full TD-DFT calculation is still an open question.

It is believed that the response of the continuum electrons to the perturbation will be small compared to the response of the bound states due to the lesser importance of electron correlations between the nonlocal continuum electrons, but this concept is not well studied in the context of photo-absorption calculations using TD-DFT. In this work we ignore the dynamic response of the continuum electrons by limiting the sum in equation \ref{ind_chi_gf} to be over indices corresponding only to bound KS states, i.e. our calculation ostensibly limits the photo-absorption cross-section to include bound-bound and bound-free transitions. A consequence of this choice is that only the bound-bound and bound-free absorption channels are allowed to mix. There is a plausibility argument presented in Appendix A as to why the dynamic response of free electrons is negligible in the context of the results presented in this paper.

\section{Results}
\subsection{Verification of Code Functionality}
Here we present a series of plots that show agreement with previous applications of TD-DFT to the study of photo-absorption. These comparisons are intended to show that our TD-DFT code and use of the \texttt{Tartarus} ground state is able to recover accepted results, including in the low-temperature limit. 

\begin{figure}[h]
\centering
\includegraphics[scale=0.35]{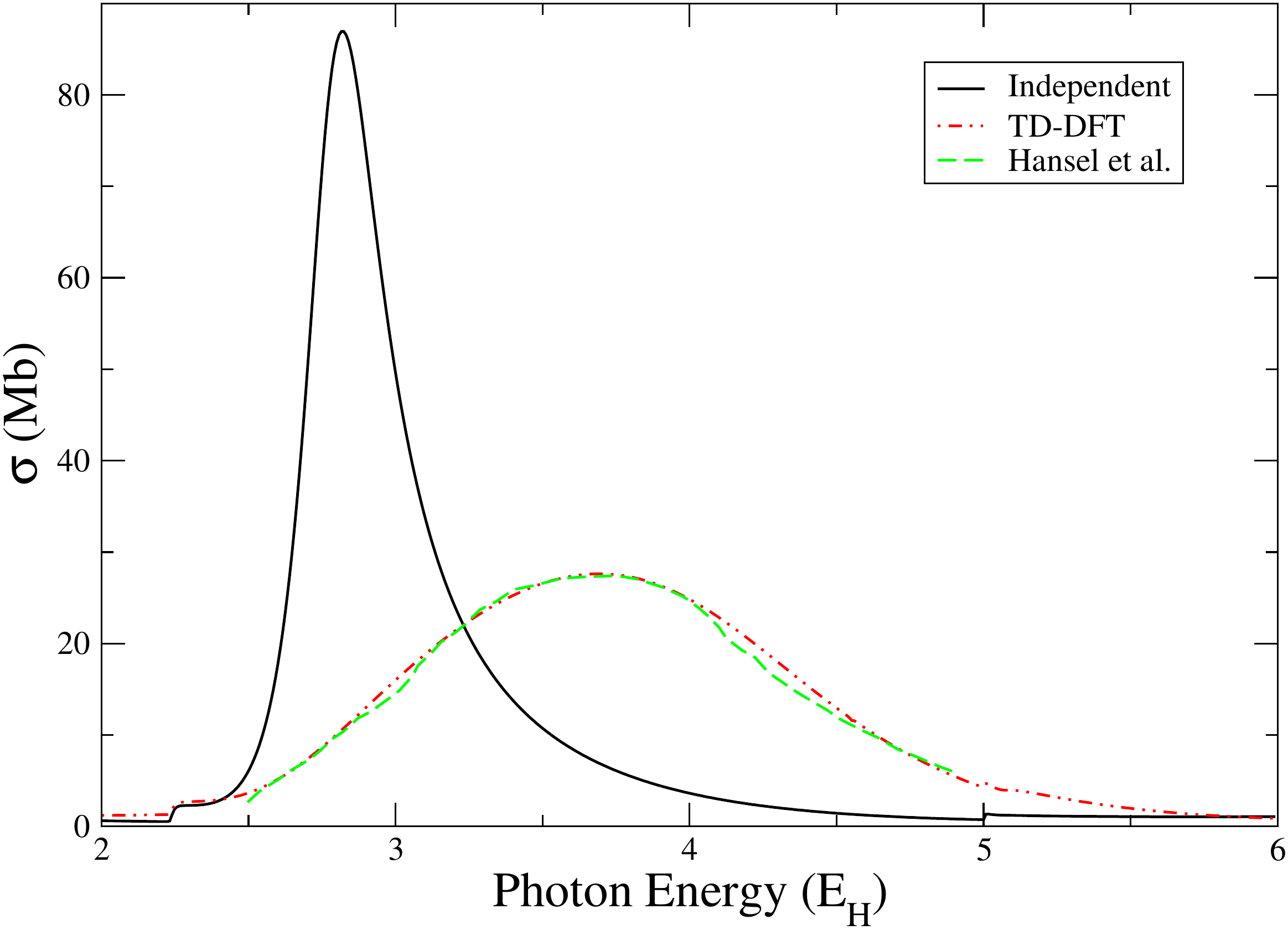}
\caption{The photo-absorption cross-section for a neutral, cold xenon gas is shown. The independent particle results are shown in solid black, the TD-DFT results are in dash-dot red, and experimental data from Hansel et al. \cite{hansel65} are in dashed green. These results indicate that our finite-temperature TD-DFT code can recover the accepted, zero-temperature limit using a \texttt{Tartarus} electronic ground state.}
\label{xeno_ZS}
\end{figure}

Figure \ref{xeno_ZS} shows that the code is able to recover the zero-temperature, isolated atom limit for xenon gas, as was a major result of ZS's original paper \cite{zangwill80}. The independent particle results (i.e. those utilizing $\chi_{0}$ in place of $\chi$) are labeled with ``Independent'' and the linear response TD-DFT results are labeled ``TD-DFT''. This convention will be maintained throughout this paper. The TD-DFT cross-section shows a reduction and redistribution of the bound-free absorption peak towards higher frequencies. At these conditions for xenon, the only allowed bound-bound transitions in the dipole approximation are between fully occupied states, which consequently excludes such transitions. Thus the only features present in the cross-section are from bound-free absorption channels. 

\begin{figure}[h]
\centering
\includegraphics[scale=0.35]{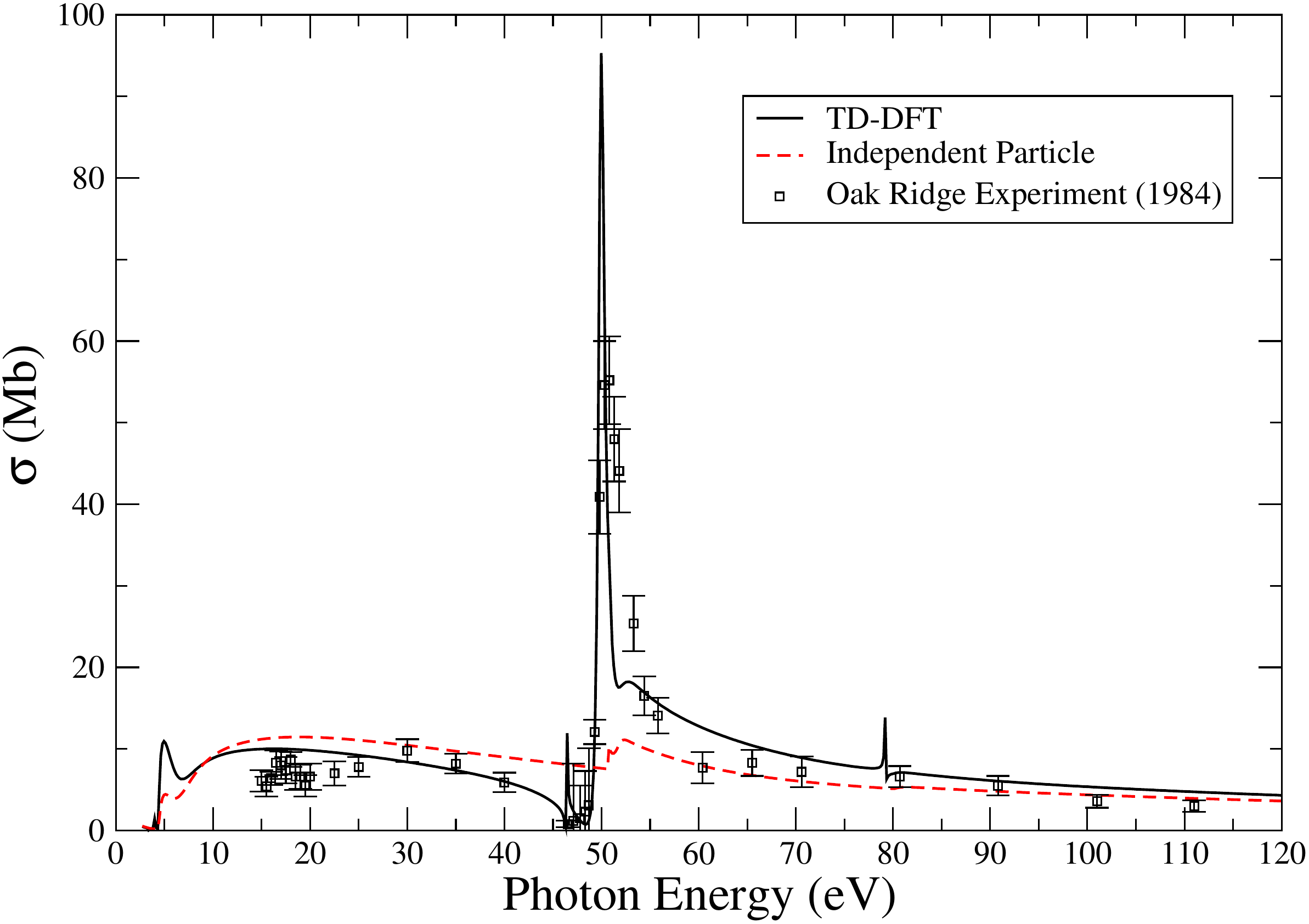}
\caption{The photo-absorption cross-section for neutral, cold manganese gas is shown. The TD-DFT results are shown in solid black and are compared to experimental results taken at Oak Ridge National Laboratory \cite{oak_ridge84}. The Fano resonance apparent in the TD-DFT results align very well with the experimental data, and the independent particle results in dashed red show that the channel mixing is necessary to reproduce the behavior seen from the experiment. This case was also studied by Liberman et al. \cite{liberman87} where similar results were obtained.}
\label{mang_cold}
\end{figure}

The significant redistribution of the cross-section seen in figure \ref{xeno_ZS} is due to the so-called channel mixing effect. This effect arises when the electrons respond to each other after a perturbation. These inter-electron interactions lead to screening and anti-screening of various absorption processes. This concept can equivalently be thought of as interference between previously independent excitation channels, e.g. a 4d photoionization channel interfering with a 4p photoionization channel. In essence, this effect arises when we account for additional correlations between electrons during some dynamic event.

Figure \ref{mang_cold} shows the effect of channel mixing through the emergence of Fano resonances in the photo-absorption cross-section for neutral, cold manganese. The large feature near $50$ $\textrm{eV}$ comes from a 3p-3d bound-bound resonance interfering with the bound-free background. This feature compares well with experimental data taken at Oak Ridge National Laboratory \cite{oak_ridge84}, demonstrating that the interference pattern is prominent for some systems under certain conditions. It should be noted here that the Fano resonances appear around allowed bound-bound transitions, but given that only the total cross-section is obtainable via TD-DFT, the bound-bound feature is embedded in this resonance structure (i.e. TD-DFT does not allow us to formally separate the spectra into bound-bound, bound-free, and free-free parts). In this way the Fano resonance could be qualitatively thought of as a modified bound-bound feature, with the line-shape due to correlations with other perturbed electrons (i.e. absorption channel interference). 

\begin{figure}[h]
\centering
\includegraphics[scale=0.35]{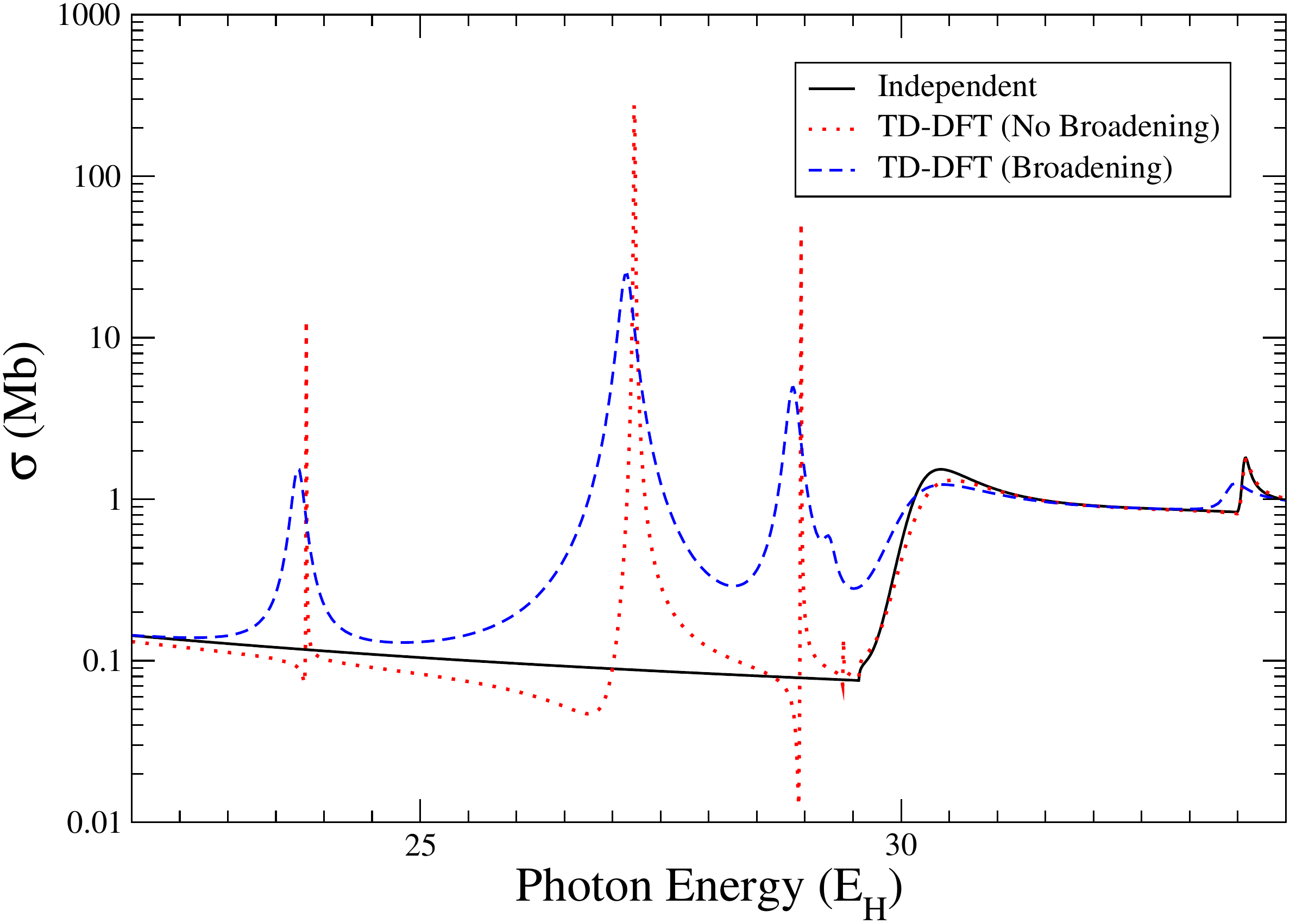}
\caption{The photo-absorption cross-section for aluminum at a temperature of $100$ $\textrm{eV}$ and solid density is shown. The independent particle results are in solid black, and the TD-DFT results with broadening are shown in dashed blue. The same TD-DFT calculation, but with no broadening (i.e. no imaginary factor is added to the state energies), is shown in red dots. The no-broadening results compare well to similar work by Grimaldi et al. \cite{grimauldi85}. These results demonstrate that the channel mixing effect gives a Fano-type profile centered around the bound-bound excitation energies, and effectively serve to broaden the bound-bound cross-section by such a profile.}
\label{iron_grim_comp}
\end{figure}

The same Fano-like features are shown in figure \ref{iron_grim_comp} for an iron plasma at $100$ $\textrm{eV}$ and solid density. This case was studied in reference \cite{grimauldi85}, and here we replicate the results with and without a rough broadening model applied. Without broadening, it is clear that the interference pattern shows the distinctive asymmetric profile expected from the channel mixing technique. When the simple Stark broadening model is applied, the asymmetry around the bound-bound excitation energies is not visible in the cross-section. This is because the bound-bound transitions no longer occur at fixed excitation energies due to the level widths applied through the imaginary part of the excitation energy. As such, the cross-section is smoothed out such that the asymmetry is no longer visible. It should also be noted here that the broadening model has little effect on the bound-free tails of the cross-section (e.g. at frequencies near $32$ $\mathrm{E_{H}}$). 

\subsection{Results Relevant to Sandia Experiments}
The under-predicted opacities obtained from Sandia's Z-machine experiments \cite{bailey2015} have provided a strong impetus for reviewing the physics employed in theoretical opacity modeling, which naturally leads to the question of whether channel mixing, as incorporated in linear response TD-DFT, can explain the missing opacity. It is clear from looking at the frequency range around $4$ $\mathrm{E_{H}}$ in figure \ref{xeno_ZS} that channel mixing can lead to significant redistribution and increase in the cross-section over certain frequency ranges, with a corresponding decrease in others. 

\begin{figure}[h]
\centering
\includegraphics[scale=0.35]{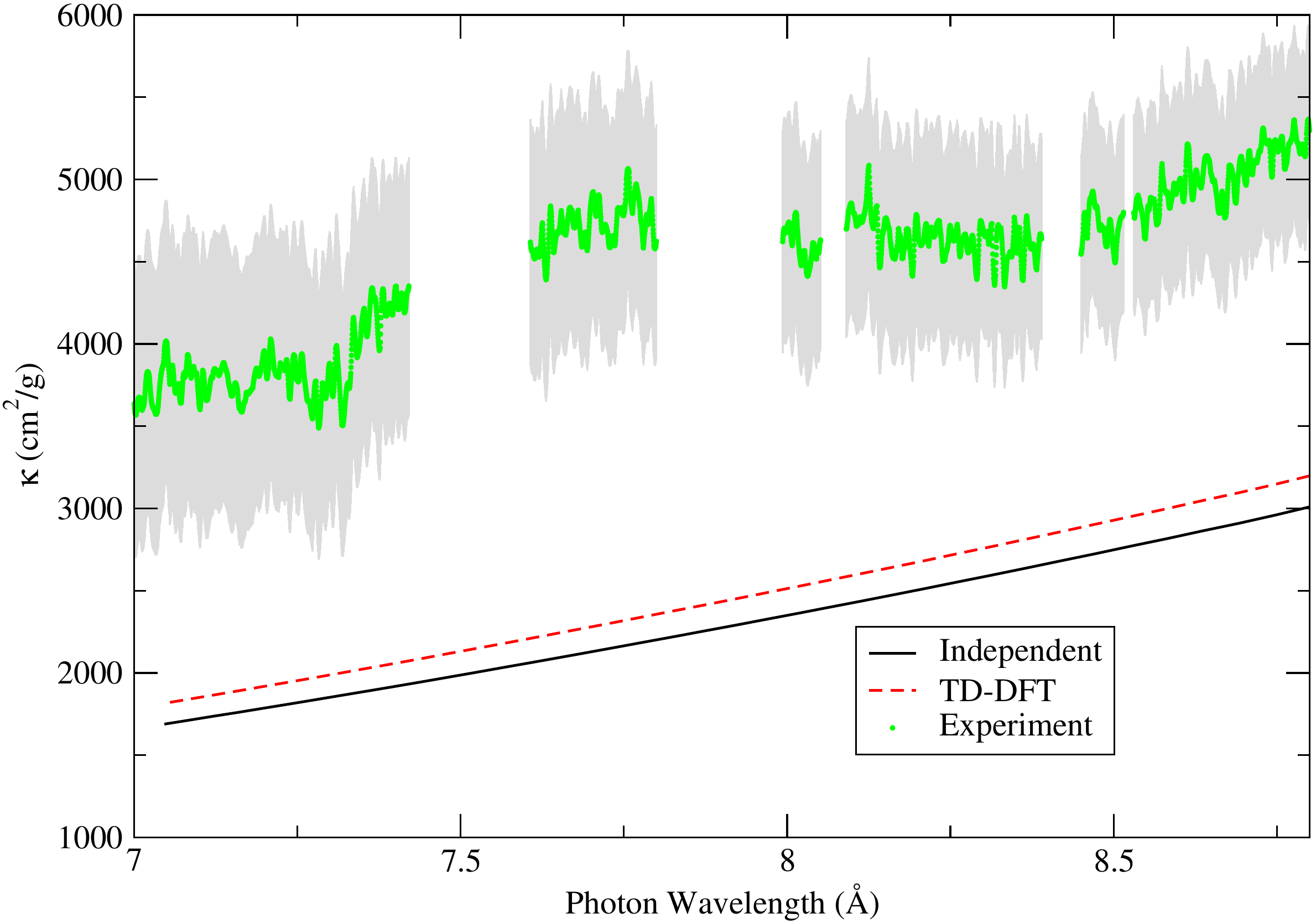}
\caption{The opacity for iron at a temperature of $170$ $\textrm{eV}$ and $0.1$ $\textrm{g}/\textrm{cm}^{3}$ is shown compared to the data from the Sandia experiment at $170$ $\textrm{eV}$. The comparison is shown only in the range of bound-free absorption, where the average atom model should give more quantitatively relevant results. The experiment is shown in solid green with error bars, the independent particle results are shown in solid black, and the TD-DFT results are shown in dashed red.}
\label{expt_comp}
\end{figure}

\begin{figure}[h]
\centering
\includegraphics[scale=0.35]{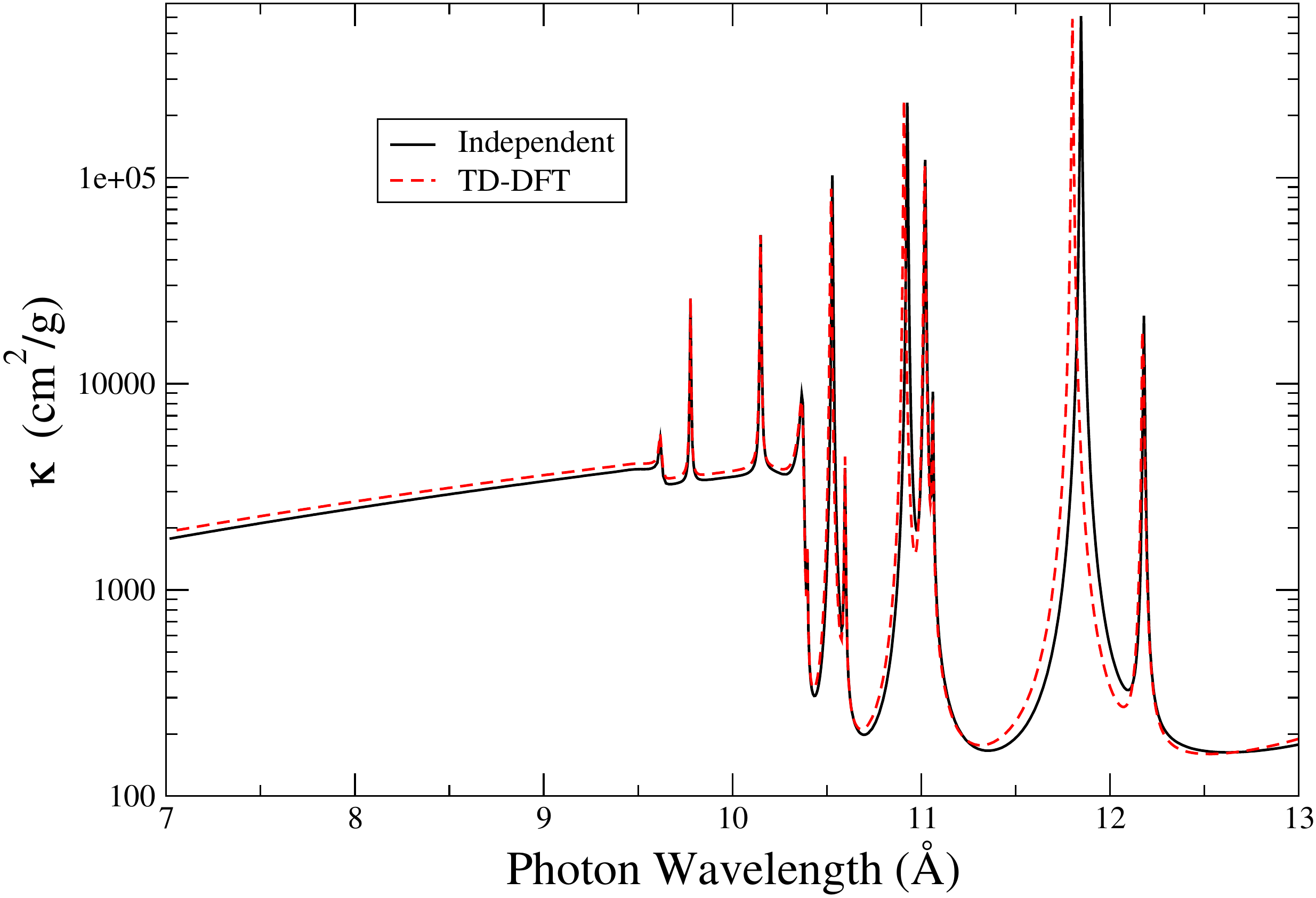}
\caption{The opacity for iron at a temperature of $175$ $\textrm{eV}$ and $0.4$ $\textrm{g}/\textrm{cm}^{3}$ is shown. These conditions are close to those of the Sandia experiment, and the difference between the TD-DFT results (dashed red) and the independent particle results (solid black) demonstrate that channel mixing only contributes an increase in the opacity of about $8$ in the $7-9.5$ ${\mathrm{\AA}}$ range.}
\label{iron_opac}
\end{figure}

\begin{figure}[h]
\centering
\includegraphics[scale=0.35]{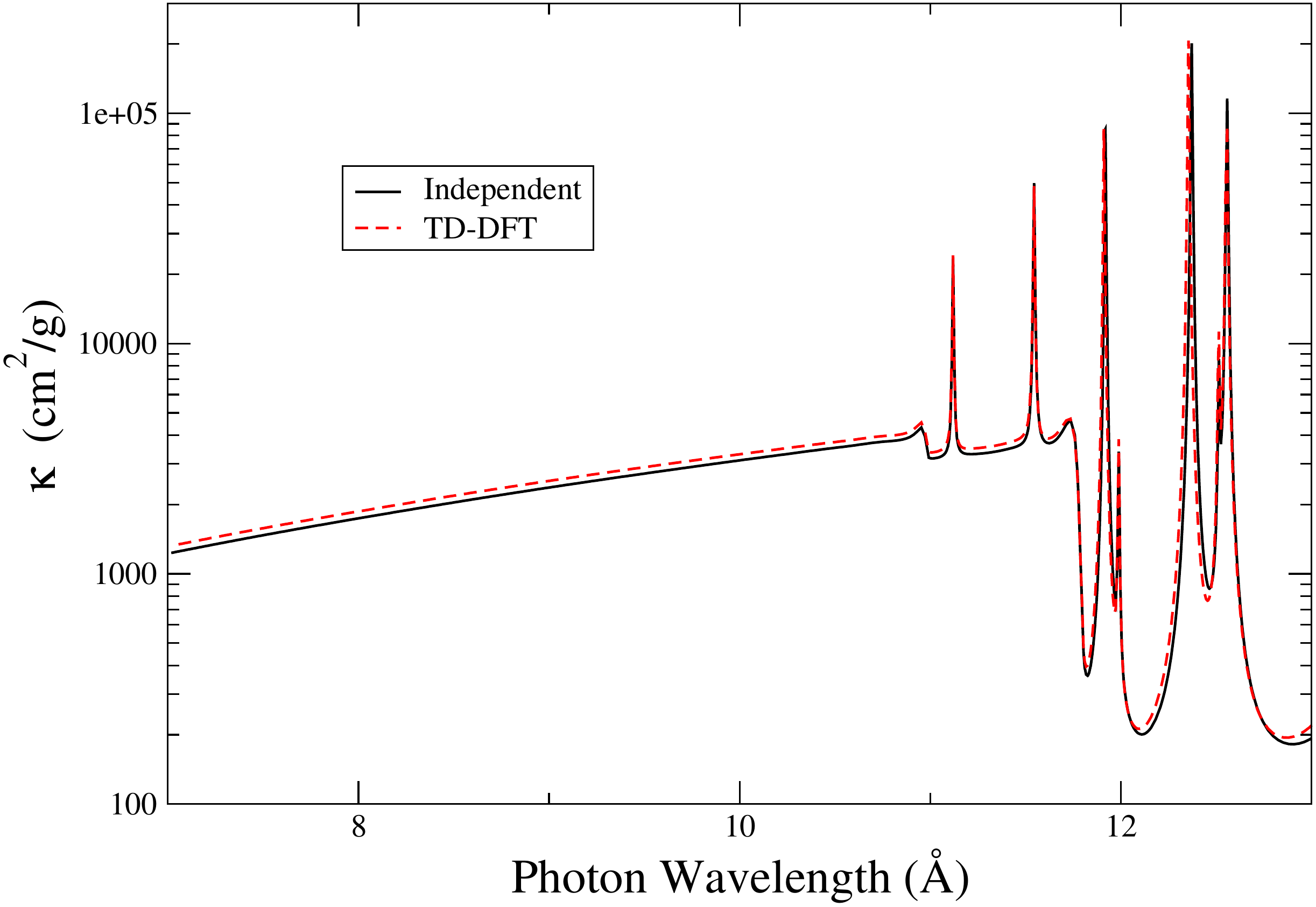}
\caption{The opacity for chromium at a temperature of $175$~$\textrm{eV}$ and $0.4$ $\textrm{g}/\textrm{cm}^{3}$ is shown. The results for the TD-DFT (dashed red) and independent particle (solid black) results do not show any change in trend from what is displayed in the iron calculation.}
\label{chromium_opac}
\end{figure}

\begin{figure}[h]
\centering
\includegraphics[scale=0.35]{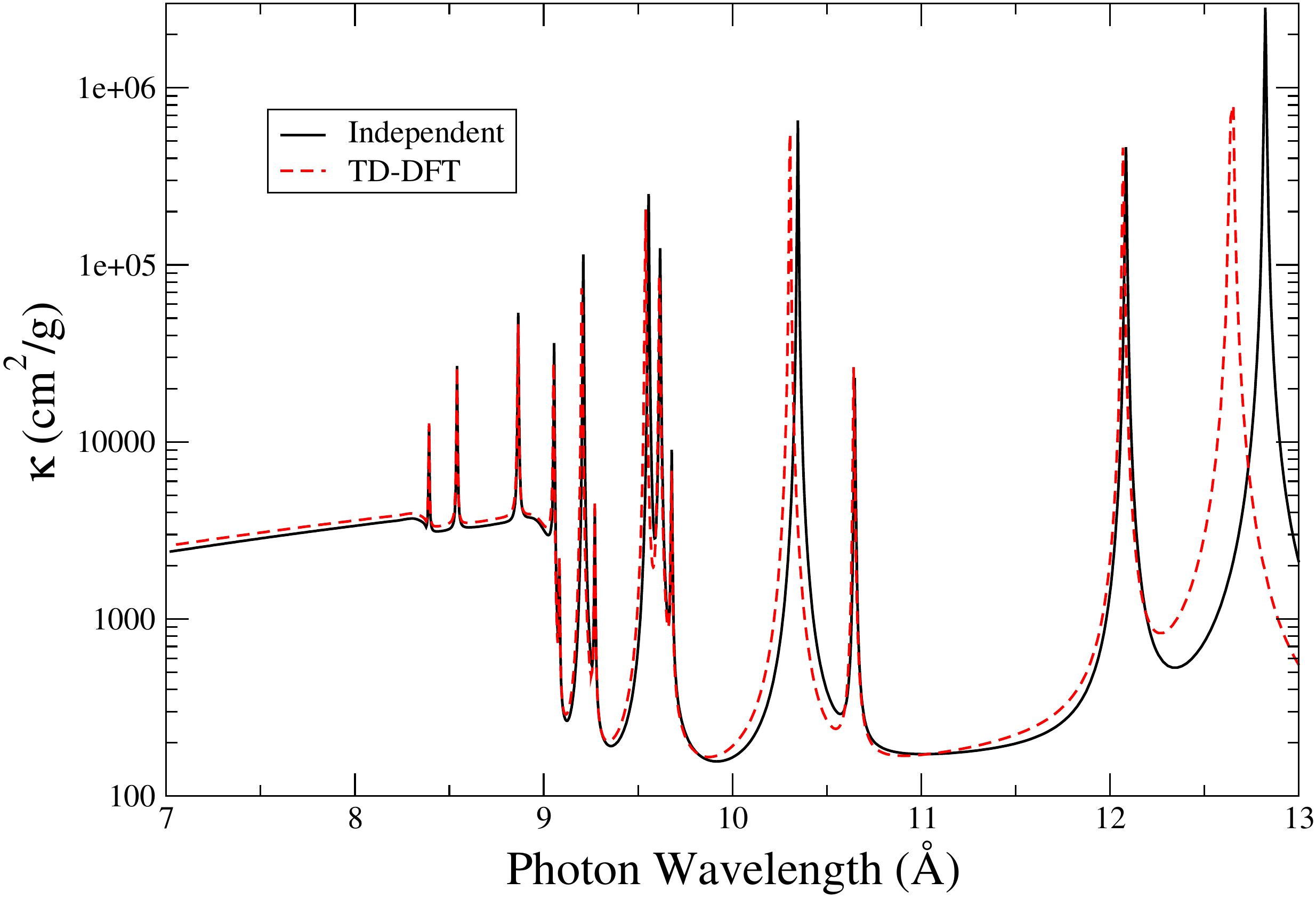}
\caption{The opacity for nickel at a temperature of $175$ $\textrm{eV}$ and $0.4$ $\textrm{g}/\textrm{cm}^{3}$ is shown. The results for the TD-DFT (dashed red) and independent particle (solid black) calculations do not show any change in trend from what is displayed in the iron calculation.}
\label{nickel_opac}
\end{figure}

In order to determine if similar trends would occur for iron, nickel, and chromium, we calculated the TD-DFT cross-section over a wide range over plasma conditions for each element. Though direct comparison of bound-bound features to the experimental data is not useful due to our use of an average atom physical model for the plasma, we can still examine the qualitative effects of channel mixing by comparing the TD-DFT results to the independent particle calculation in areas of the spectrum dominated by bound-free absorption. In figure \ref{expt_comp}, we show a comparison of our calculation to the Sandia experiment \cite{bailey2007,bailey2015} results for iron at $170$ $\textrm{eV}$ in a range of wavelengths corresponding to predominantly bound-free absorption. The increase in the opacity seen in the TD-DFT results due to channel mixing do not rectify the discrepancy between the theoretical models and the experimental results.

We show the opacities for iron, chromium, and nickel at $175$ $\textrm{eV}$, $0.4$ $\textrm{g}/\textrm{cm}^{3}$ in figures \ref{iron_opac}, \ref{chromium_opac}, and \ref{nickel_opac}, respectively. The same broadening model is applied to both the independent particle and TD-DFT results in each calculation, but it is important to note that calculations without the broadening applied do not display any qualitative differences in the bound-free tail between 7--9.5 ${\mathrm{\AA}}$ in each case. In each case, we see similar qualitative trends. The bound-bound features in the TD-DFT calculation are further broadened, shifted towards higher energies, and peak at smaller values compared to their independent particle counterparts. This redistribution leads to an increase in the bound-free tail of the cross-section that was shown to be under-predicted in some of the experiments. The increase is relatively small, with the TD-DFT results predicting around an $8$$\%$ increase relative to the independent particle calculations. This increase is small compared to the discrepancy between theory and experiment, which is greater than ~50$\%$ in this wavelength range. 

\begin{figure}[h]
\centering
\includegraphics[scale=0.35]{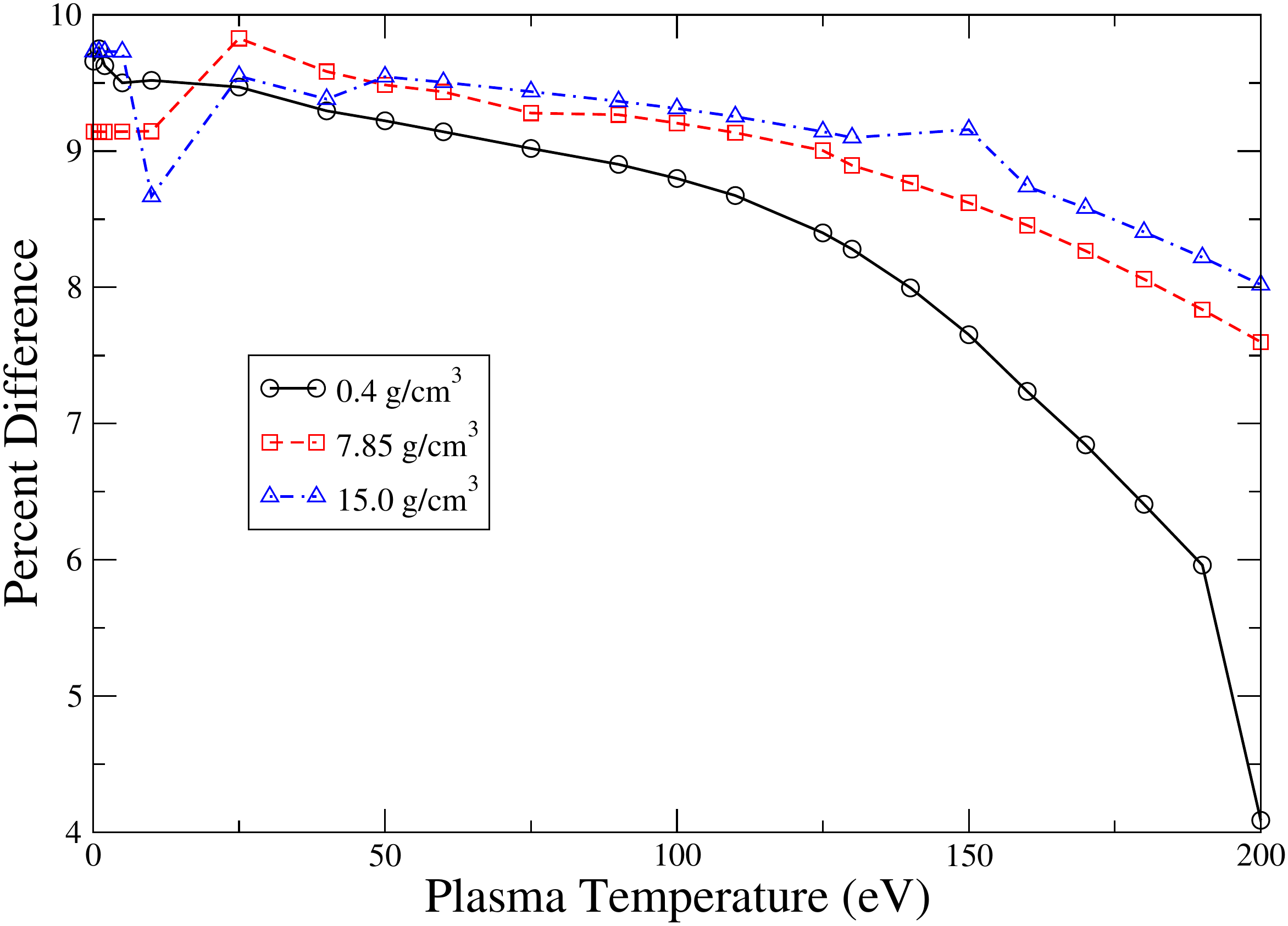}
\caption{The percent difference between the TD-DFT and independent particle results versus temperature in iron is shown at a fixed wavelength ($\lambda=9.1$ $\mathrm{\AA}$). The solid black line corresponds to a plasma density of $0.4$ $\textrm{g}/\textrm{cm}^{3}$, the dashed red line corresponds to a plasma density of $7.85$ $\textrm{g}/\textrm{cm}^{3}$ (solid density), and the dash-dot blue line corresponds to a plasma density of $15.0$ $\textrm{g}/\textrm{cm}^{3}$. The general trend shows that the effect of channel mixing on this part of the bound-free cross-section reduces as the plasma temperature increases. The same trend is seen at higher densities, though the rate of decrease is muted as the density increases. The same analysis of chromium and nickel does not show any qualitative differences in the temperature and density trends for those elements. The lines in this figure are meant to serve only as guides to the eye and do not represent an accurate interpolation between different temperatures.} 
\label{perc_diff_iron}
\end{figure}

The trend that we observe does not change in any qualitatively unexpected ways for each element or at different plasma conditions. The percent difference between the TD-DFT and independent particle results at a fixed wavelength ($\lambda = 9.1$ $\mathrm{\AA}$) for iron is shown in figure \ref{perc_diff_iron} at three different densities. This figure shows that there are not any particular plasma conditions for which the TD-DFT calculation predicts significantly higher opacities than the independent particle calculation. The same general temperature and density dependence seen in figure \ref{perc_diff_iron} was observed for the chromium and nickel calculations. 

From figure \ref{perc_diff_iron}, we can see that, as the temperature of the plasma increases, the overall effect of channel mixing decreases. At higher temperatures, the electrons in the plasma exhibit weaker correlations with each other, consequently weakening the effect of channel mixing. We also see that the decrease in the effect due to temperature softens as the density increases. This behavior is due to the plasma becoming more degenerate at higher densities, leading to stronger electron correlations and consequently stronger channel mixing effects.

\section{Conclusions}
We have shown that the \texttt{Tartarus} average atom model provides a suitable ground state density for the application of linear response TD-DFT to the study of photo-excitation in plasmas. The channel mixing effects that arise with TD-DFT can show significant redistribution of photo-absorption cross-sections, as well as the emergence of Fano resonances around bound-bound excitation energies in the cross-section. These results compare well to similar calculations done in the past and to experimental data of cold, neutral gases.

When applying the linear response TD-DFT calculation to plasmas motivated by opacity experiments carried out at Sandia National Laboratories, we see that the channel mixing process generally leads to an increase in the bound-free cross-sections, but the magnitude of the effect is not sufficient to explain the discrepancy between theory and experiments. Further, there is no significant change in the expected trends at different plasma densities, temperatures, and compositions. The overall changes in cross-sections due to incorporating channel mixing appear to be most significant for lower temperature plasmas. The decrease in effect with higher temperatures is lessened for high density plasmas due to the higher electron degeneracy in such systems. Overall it appears that channel mixing is an effect that can contribute changes on the order of a $10$$\%$ over certain frequency ranges in bound-free cross-sections, with possibly more significant changes in lower temperature systems. 

\section*{Acknowledgments}
We thank T. Nagayama for providing the experimental data. This work was performed under the auspices of the United States Department of Energy under contract DE-AC52-06NA25396.

\section{Appendix A}
Linear response TD-DFT does not formally allow us to separate the cross-sections into bound-bound, bound-free, and free-free components, as is commonly done in most other photo-absorption models. In this sense, each of the cross-sections shown in this work are representative of the total photo-absorption cross-section without considering excitations of continuum electrons. This means that these cross-sections are most comparable to the sum of bound-bound and bound-free cross-sections when comparing to other models. However, since it is not formally possible to separate out the contribution of the response of free electrons to each other and the bound electrons in the system, it is reasonable to consider whether or not the exclusion of such electrons could lead to significant or misleading results in the calculated cross-sections. 

In order to quantify the effect of ignoring certain electrons in the calculation, we look at the cross-sections calculated using the TD-DFT formalism, but with certain KS states excluded from the calculation of $\chi_{0}$. Here we will present a formulation that we believe justifies ignoring the response of continuum electrons in the results presented in this work. First we define a truncated KS response function:
\begin{equation}
\label{trunc_resp}
\begin{split}
\chi_{0}^{C}(\vec{r},\vec{r'},\omega) =& \sum_{i \in C} f_{i}\psi_{i}^{*}(\vec{r})\psi_{i}(\vec{r'})G(\vec{r},\vec{r'},E_{i}+ \omega) +\\ &\sum_{i \in C}f_{i}\psi_{i}(\vec{r})\psi_{i}^{*}(\vec{r'})G^{*}(\vec{r},\vec{r'},E_{i}-\omega)
\end{split}
\end{equation}
where $C$ defines a subset of the ground state KS states. For example, for the calculations presented previously in this work, we use a definition of $C$ such that it contains all KS states with $E_{i}<0$. For simplicity, we omit labelling such a set, and, in our present work, $\chi_{0}$ is assumed to only contain contributions from bound KS states and is referred to as the ``full'' response function. We subtract a partial response function from a ``full'' response function to obtain
\begin{equation}
\label{delta_resp}
\Delta \chi_{0}^{C} = \chi_{0} - \chi_{0}^{C}
\end{equation}

Using $\Delta \chi_{0}^{C}$ in the equations described in Sections 2.1 and 2.2 yields a photo-absorption cross-section that approximates the response of the system without participation of the electrons in set $C$. We label such a cross-section as $\sigma^{C}$. The contribution to a cross-section due to channel mixing can be seen by subtracting the independent cross-section from the TD-DFT cross-section, i.e. $\sigma_{LR}-\sigma_{0}$, where the $LR$ label indicates the results from TD-DFT and the $0$ label indicates the independent particle results. 

Finally, in order to quantify the approximate contribution to the channel mixing effects that a set of states, $C$, provides, we look at the differences between the channel mixing contribution of the ``full'' cross-section and a partial cross-section, i.e.
\begin{equation}
\label{delta_delta_sigma}
\Delta \sigma^{C} = (\sigma_{LR}-\sigma_{0}) - (\sigma_{LR}^{C}-\sigma_{0}^{C})
\end{equation}
The quantity $\Delta \sigma^{C}$ approximates the channel mixing contribution to the TD-DFT cross-section due to states in set C. If $\Delta \sigma^{C}$ is a small quantity, then ignoring the interference of absorption processes associated with states in set C (i.e. ignoring absorption channels where states in set C are the ``initial'' or ``final'' state) will have little impact on the total TD-DFT cross-section.
\begin{figure}[h]
\centering
\includegraphics[scale=0.35]{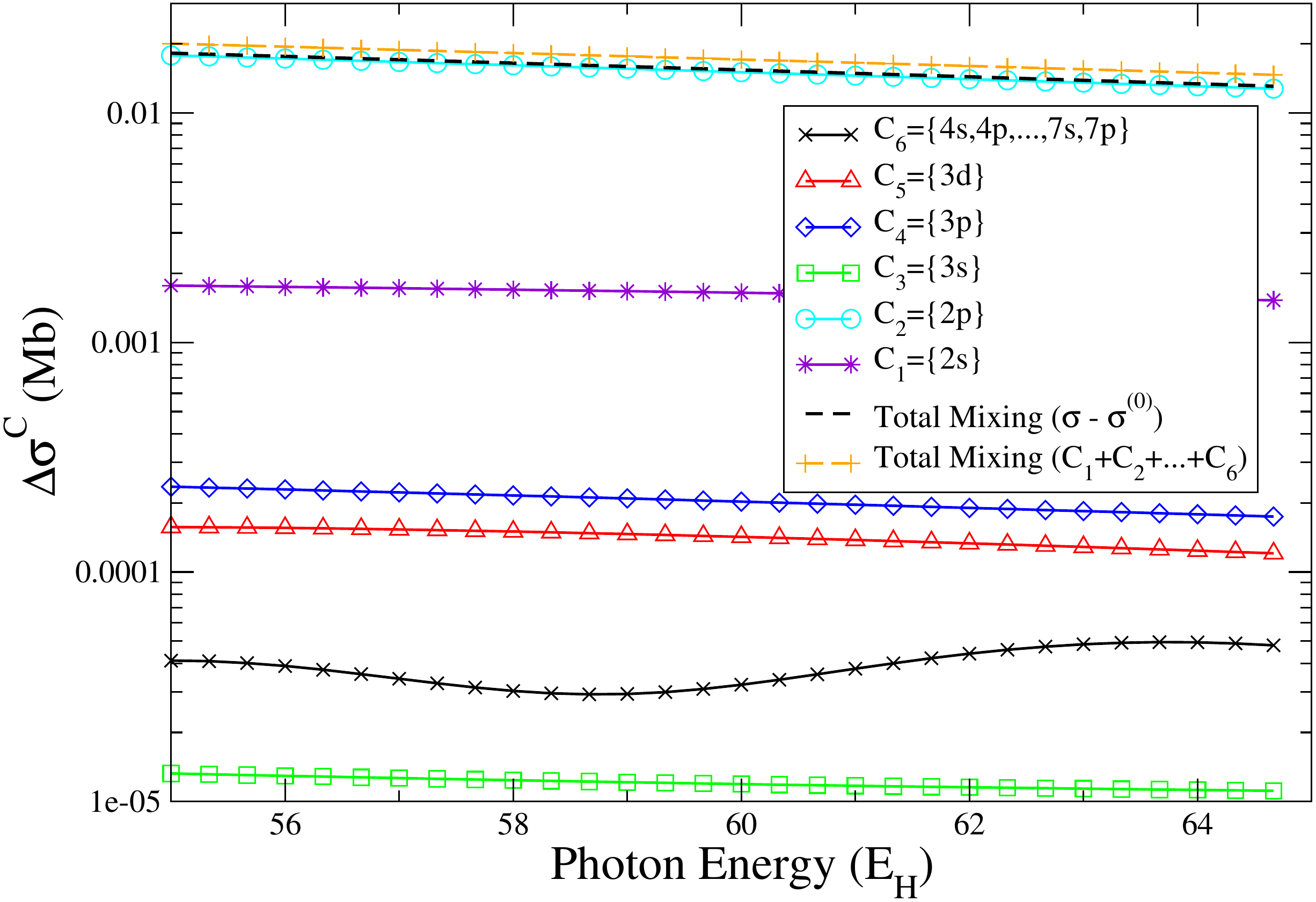}
\caption{The relative contribution to the channel mixing effect versus frequency for various sets, $C$, of single-particle KS states is shown. The more weakly bound states are compiled into set $C_{6}$, and their contribution is very small relative to the most strongly contributing states -- the $2s$ ($C_{1}$) and $2p$ ($C_{2}$). This procedure indicates that the continuum electrons can be ignored in the calculation of the response functions without introducing any significant error into the resulting cross-sections.}
\label{delta_sig_ctm_arg}
\end{figure}

For iron at $185$$\textrm{eV}$ and $0.4$$\textrm{g}/\textrm{cm}^{3}$, we calculated $\Delta \sigma^{C}$ for various choices of $C$. The results are shown in figure \ref{delta_sig_ctm_arg}, from which we can see that the strongest contributions to the channel mixing effect come from the 2p absorption channels, i.e. $C_{2}$. This is to be expected because, at the range of photon energies in question, the photo-absorption cross-section is mostly due to the $2s$ and $2p$ photoionization features, and therefore the $2p$ absorption channel interfering with the $2s$ absorption yields the largest difference to the cross-section. The next highest contribution comes from the $2s$ absorption channel interfering with the $2p$ absorption. The contribution from the $3s$ state is the smallest shown, despite being relatively close in energy to the $2s$ and $2p$ states. This behavior is due to the photo-absorption cross-section associated with the $3s$ being very small in this frequency range, and therefore the channel mixes very poorly with the more prominent absorption channels.

The most weakly bound states (represented by the set $C_{6}$) contribute very little difference to the cross-section relative to most other absorption channels. These states are the closest in energy to the occupied continuum states and, due to being weakly bound, they are also most similar in terms of spatial extent. Given how little these states contribute to the cross-section differences, it is reasonable to assume that the contribution of the continuum electrons would likewise be quite small. 

As emphasized before, $\Delta \sigma^{C}$ is just an approximation to the contribution from channel mixing due to set $C$. Although the total response function, $\chi_{0}$, can be represented as the sum of truncated response functions, $\chi_{0}^{C}$, without any approximation, the same cannot be said of the total cross-section difference and the $\Delta \sigma^{C}$ due to the self-consistent scheme used to determine the cross-sections. As we can see from the line labeled Total Mixing $(C_{1} + C_{2} + ... + C_{6})$ in figure \ref{delta_sig_ctm_arg}, adding up the truncated contributions over-predicts the total mixing contribution. This behavior indicates that the approximation utilized to determine $\Delta \sigma^{C}$ has a tendency to over-predict the contribution from sets of states. This result further supports the argument that, if we can ignore a large set of weakly bound electrons for these plasma conditions, we can ignore the contributions of continuum electrons. This argument should hold true even if there are continuum resonances, as is often the case for partially ionized, dense plasmas.

\section*{References}
\bibliographystyle{unsrt}
\bibliography{phys}

\end{document}